\documentclass[conference,9pt]{IEEEtran}
\IEEEoverridecommandlockouts
\usepackage{amssymb}
\usepackage{amsmath}
\usepackage{booktabs}
\usepackage{stfloats}
\usepackage{tikz}
\usetikzlibrary{decorations.pathreplacing}
\usepackage{arydshln}
\usepackage{enumitem}
\usepackage{multirow}
\usepackage{epsfig}
\usepackage{multicol,lipsum,xparse}
\usepackage{caption}
\usepackage{subcaption}
\usepackage{stmaryrd}
\usepackage{hyperref}
\usepackage{tikz}
\usepackage{algorithm}
\usepackage{algorithmicx}
\usepackage{algpseudocode}
\usepackage{pifont}
\usepackage{cleveref}

\usepackage{todonotes}
\usepackage{cite}
\usepackage{amsmath,amssymb,amsfonts}
\usepackage{graphicx}
\usepackage{textcomp}

\usepackage{adjustbox}
\usepackage{xcolor}
\usepackage[a4paper, total={184mm,239mm}]{geometry}
\def\BibTeX{{\rm B\kern-.05em{\sc i\kern-.025em b}\kern-.08em
    T\kern-.1667em\lower.7ex\hbox{E}\kern-.125emX}}

\begin{document}
\pagenumbering{roman}

\title{BoolE: Exact Symbolic Reasoning via Boolean Equality Saturation}

\newcommand{\fixme}[1]{\textcolor{red}{\small [~#1~]}}

\author{\IEEEauthorblockN{Jiaqi Yin*, Zhan Song*, Chen Chen, Qihao Hu, Cunxi Yu}
\IEEEauthorblockA{
\textit{University of Maryland, College Park}\\
College Park, US \\
\{jyin629,cunxiyu\}@umd.edu}

\thanks{*: co-first authors. J. Yin and C. Yu contributed the idea. J. Yin implemented the framework. Z. Song conducted experimental and baseline benchmarking setups. All authors contributed to the writing.}
}

\newcommand*\circled[1]{\raisebox{.4pt}
                    {\tikz[baseline=(char.base)]{
            \node[shape=circle,draw,inner sep=1pt, style={fill=white, text=black}, scale=0.75] (char) {\textbf{#1}};}}}
\maketitle

\begin{abstract}

Boolean symbolic reasoning for gate-level netlists is a critical step in verification, logic and datapath synthesis, and hardware security. Specifically, reasoning datapath and adder tree in bit-blasted Boolean networks is particularly crucial for verification and synthesis, and challenging.
Conventional approaches either fail to accurately (exactly) identify the function blocks of the designs in gate-level netlist with structural hashing and symbolic propagation,
or their reasoning performance is highly sensitive to structure modifications caused by technology mapping or logic optimization.
This paper introduces \textbf{BoolE}, an exact symbolic reasoning framework for Boolean netlists using equality saturation. 
BoolE optimizes scalability and performance by integrating domain-specific Boolean ruleset for term rewriting.
We incorporate a novel extraction algorithm into BoolE to enhance its structural insight and computational efficiency, which adeptly identifies and captures multi-input, multi-output high-level structures (e.g., full adder) in the reconstructed e-graph.

Our experiments show that BoolE surpasses state-of-the-art symbolic reasoning baselines, including the conventional functional approach (ABC) 
and machine learning-based method (Gamora). Specifically, we evaluated its performance on various multiplier architecture with different configurations. 
Our results show that BoolE identifies $3.53\times$ and $3.01\times$ more exact full adders than ABC in carry-save array and Booth-encoded multipliers, respectively.
Additionally, we integrated BoolE into multiplier formal verification tasks, where it significantly accelerates the performance of traditional formal verification tools using computer algebra, demonstrated over four orders of magnitude runtime improvements.



\end{abstract}

\section{Introduction}\label{sec:intro}

Boolean symbolic reasoning, which extracts word-level abstractions from gate-level netlists, plays a critical role in electronic design automation (EDA) such as logic synthesis, datapath optimization, and formal verification \cite{ciesielski2015verification, mahzoon2019revsca, mahzoon2021revsca}. Additionally, the globalization of VLSI design and manufacturing processes has amplified the need for detecting malicious logic, such as hardware Trojans, to ensure hardware security \cite{meade2016gate, rajendran2021novel}. However, conventional approaches, such as those based on cut enumeration \cite{pan1998new, brayton2010abc, subramanyan2013reverse}, structural hashing \cite{yu2017fast, mahzoon2019revsca, mahzoon2021revsca}, and machine learning (ML) techniques \cite{alrahis2021gnn, wang2022efficient, wu2023gamora, deng2024less}, face several critical limitations when addressing complex reasoning tasks, particularly in \textit{bit-blasted} non-linear arithmetic netlists:


\textbf{(1)} \textit{Static structural limitation} -- Structural approaches, such as those based on cut enumeration and local structural hashing \cite{brayton2010abc, pan1998new, subramanyan2013reverse}, often use circuit topology for shape hashing to identify structurally similar wires and form word-level abstractions. Alternatively, they rely on reference libraries to map sub-circuits by matching local truth tables \cite{yu2017fast, cakir2018reverse}. In these cases, the original high-level structures of a given netlist—such as carry-chains and adder trees—are often fragmented or altered during heavy logic optimization and technology mapping, rendering them unrecognizable. 
\textbf{(2)} \textit{Lack of completeness and exactness}: State-of-the-art (SOTA) Boolean methods for symbolic reasoning focus on detecting Negation-Permutation-Negation (NPN) classes of functional components, which represent groups of structurally similar but not exact functional blocks. While these abstractions are useful for technology mapping \cite{huang2013fast}, they fail to provide exactness guarantees, which are critical for real-world applications like formal verification. Furthermore, ML-based approaches, such as Gamora \cite{wu2023gamora}, HOGA \cite{deng2024less}, and DeepGate \cite{li2022deepgate, shi2024deepgate3}, lack completeness and correctness guarantees due to the inherently probabilistic nature of ML models. 
\textbf{(3)} \textit{Scalability challenges of functional approaches}: Functional approaches using symbolic evaluation are solver-ready (e.g., SAT-based approaches \cite{de2011satisfiability, biere2013lingeling}), but incur significant computational overhead, especially when applied to \textit{bit-blasted} non-linear arithmetic Boolean networks \cite{mahzoon2019revsca, yu2016formal, kaufmann2021amulet}.

Given the limitations of conventional approaches, we present \textbf{BoolE}, an exact symbolic reasoning framework that utilizes Boolean \textit{equality saturation}. BoolE is designed to enhance reasoning performance for complex Boolean netlists, which are technology-mapped or logic-optimized. It takes Boolean netlists as input and infers functional blocks through Boolean-level equality saturation and term rewriting. By leveraging the e-graph saturation theory \cite{tate2009equality}, BoolE offers scalable solutions to reasoning high-level blocks in heavily optimized and technology mapped arithmetic multipliers. More importantly, BoolE offers formal exactness and correctness guarantees. 
The key contributions of BoolE are summarized as follows:

\begin{enumerate}
    \item \textbf{Boolean reasoning through equality saturation:} To the best of our knowledge, BoolE is the first end-to-end framework that uses equality saturation to explore a vast space of equivalent Boolean expressions. BoolE methodologies are believed to have broader impacts beyond symbolic reasoning in Boolean domain. 

    \item \textbf{Novel exact extraction algorithm for multi-output structures:} BoolE introduces an innovative exact extraction algorithm alongside a domain-specific Boolean ruleset, capable of handling complex multi-input, multi-output high-level structures. This ensures precise and efficient logic reconstruction, with exactness and correctness guaranteed by equality saturation theory.

    \item \textbf{Comprehensive evaluation on reasoning:} Extensive experimental results demonstrate that BoolE outperforms SOTA functional, structural, and ML-based approaches by significant margins in both exact and NPN reasoning. 

    \item \textbf{End-to-end integration to real-world application:} BoolE has been seamlessly integrated with SOTA synthesis and verification tools, such as ABC, as well as formal verification backends like RevSCA2.0 \cite{mahzoon2021revsca}. A case study on the formal verification of arithmetic multipliers demonstrates over \textbf{four orders of magnitude} runtime acceleration on RevSCA-2.0.
\end{enumerate}

\section{Background}\label{sec:background}




\subsection{Boolean Networks}
Boolean networks are mathematical models that represent logical relationships between binary variables, where nodes signify Boolean variables (0 or 1) and edges denote dependencies defined by Boolean functions. These networks are fundamental in digital circuit design for modeling combinational logic, and be used in computational biology for gene regulatory networks.
An And-Inverter Graph (AIG) is a specialized Boolean network extensively used in electronic design automation (EDA). It is a directed acyclic graph (DAG) composed solely of two-input AND gates and NOT gates (inverters). AIGs offer a compact representation of Boolean functions, facilitating efficient manipulation, optimization, and verification of complex digital circuits. All combinatorial Boolean networks can be converted to AIG with De Morgan’s laws.

A cut of a Boolean network is a pair $(r, S)$, where $r$ is a node called the root, and $S$ is a set of nodes called the leaves, such that:
(1) Every path from a primary input to $r$ passes through at least one leaf in $S$.
(2) For each leaf $l \in S$, there is a path from a primary input to $r$ that goes through $l$ but no other leaf.
The size of the cut is the number of leaves $|S|$. A cut is $K$-feasible if its size $|S| \leq K$. This work focuses on 3-feasible cuts. The cut enumeration algorithm is a fundamental technique in Boolean network analysis, specifically used in symbolic reasoning and technology mapping.
By enumerating all possible $K$-feasible cuts, the cut enumeration algorithm systematically explores various sub-network configurations, enabling the detection of critical Boolean functions such as FAs and multiplexers.
The exhaustive exploration enables the identification of optimal sub-functions, enhances logic optimization, and supports technology mapping by providing a comprehensive understanding of the circuit functional structure.
Despite the effectiveness of the cut enumeration algorithm, it struggles to accommodate structural modifications such as technology mapping and logic optimization. 

NPN classification is a method used in logic synthesis and technology mapping to group Boolean functions into equivalence classes\cite{huang2013fast}. Two Boolean functions are considered NPN equivalent if one can be transformed into the other through a combination of input negations, input permutations, and output negation. Each NPN class of a representative function comprises all functions that can be derived from these transformations.

\subsection{Boolean Word-Level Block Identification}


Boolean word-level block identification in digital design aims to detect elementary functional blocks (e.g., FA) within netlists, providing critical word-level abstractions that enable logical optimization, formal verification, malicious logic detection, and other verification tasks\cite{meade2016gate, ciesielski2019understanding, mahzoon2021revsca}. Conventional approaches include structural shape hashing and functional bitslice aggregation \cite{subramanyan2013reverse,yu2017fast,li2013wordrev}, which analyze block-level netlists from structural and functional perspectives. 
Structural shape hashing assigns a unique shape to each edge in a Boolean netlist to capture its structural properties. This shape is defined as a directed graph that represents the backward-reachable logic gates from the wire.
Functional bitslice aggregation, on the other hand, relies on functional matching to group equivalent nodes and edges using the cut enumeration algorithm. 
The cut enumeration algorithm is integrated into synthesis tools like ABC. Besides, some works also focus on recover higher-level programming abstractions in hardware description language (HDL) code \cite{sisco2023loop, rao2024register}.
Recently, Graph Neural Networks (GNNs) have emerged as a promising tool for Boolean block identification~\cite{alrahis2021gnn, he2021graph, wang2022efficient, wu2023gamora, deng2024less}. Gamora~\cite{wu2023gamora}, for instance, leverages GNNs and GPU acceleration to efficiently infer high-level functional blocks from gate-level netlists.

FA is the fundamental functional block for adder trees, consisting of sum and carry signals.
which are represented by XOR and MAJ gates with identical input signals. Both the cut enumeration algorithm in ABC and the Gamora framework identify FAs based on NPN equivalence classes. 
In this paper, we refer to the logically equivalent FA as \textbf{exact FA}, and NPN equivalent FA as \textbf{NPN FA}.



\subsection{Equality Saturation}
Equality saturation\cite{tate2009equality,zhang2023better,nelson1980fast,nieuwenhuis2005proof} is a rewriting optimization technique powered by an e-graph (equivalence graph) data structure, which represents an equivalence relation over expressions. An \textbf{e-graph} $\mathcal{G}$ is formally defined as a tuple $\mathcal{G} = (V, \mathcal{R}, \lambda)$, where:

\begin{itemize}
    \item \( V \) is a finite set of vertices, referred to as \textbf{E-nodes}. Each e-node within \( V \) represents a distinct expression or sub-expression.
    \item \( \mathcal{R} \subseteq V \times V \) defines an equivalence relation on \( V \), partitioning it into equivalence classes known as \textbf{E-classes}. These E-classes are the equivalence classes generated by \( \mathcal{R} \). Nodes within the same e-class are considered equivalent under the relation \( \mathcal{R} \).
    \item \( \lambda: V \rightarrow \Sigma \times V^k \) is a labeling function that assigns each node to an operator and an ordered list of child nodes, where \( \Sigma \) represents the set of operators with \( k \) operands and \( V^k \) denotes an ordered tuple of \( k \) child nodes from \( V \).
\end{itemize}

E-graphs form the foundation of \textbf{equality saturation} optimization technique, which applies rewrite rules iteratively until no new equivalences can be introduced. The process involves: 
(1) constructing an e-graph $\mathcal{G}_0$ from the initial expression $s_0$,
(2) applying rewrite rules $(l_i, r_i)$ to incorporate equivalences $[l_i]_{\mathcal{R}} \sim [r_i]_{\mathcal{R}}$, and
(3) repeating until convergence, which indicates that no further changes can be made. E-graphs compactly store an exponential number of expressions in linear space by sharing sub-expressions, preserving all expressions and avoiding the phase-ordering problem~\cite{leverett1980overview}.
Equality saturation has been applied in numerous areas within hardware design automation~\cite{yang2021equality, ustun2022impress, coward2022automatic, chen2024syn,ustun2023equality,cai2025smoothe}. 
For our implementation, we utilize the \textbf{egg} tool~\cite{willsey2021egg}, an advanced e-graph framework, as the backend for BoolE e-graphs.

\section{Motivating Example}

\begin{figure*}[htbp]
    \centering
    \begin{subfigure}[t]{0.23\linewidth}
        \centering
        \includegraphics[width=\textwidth]{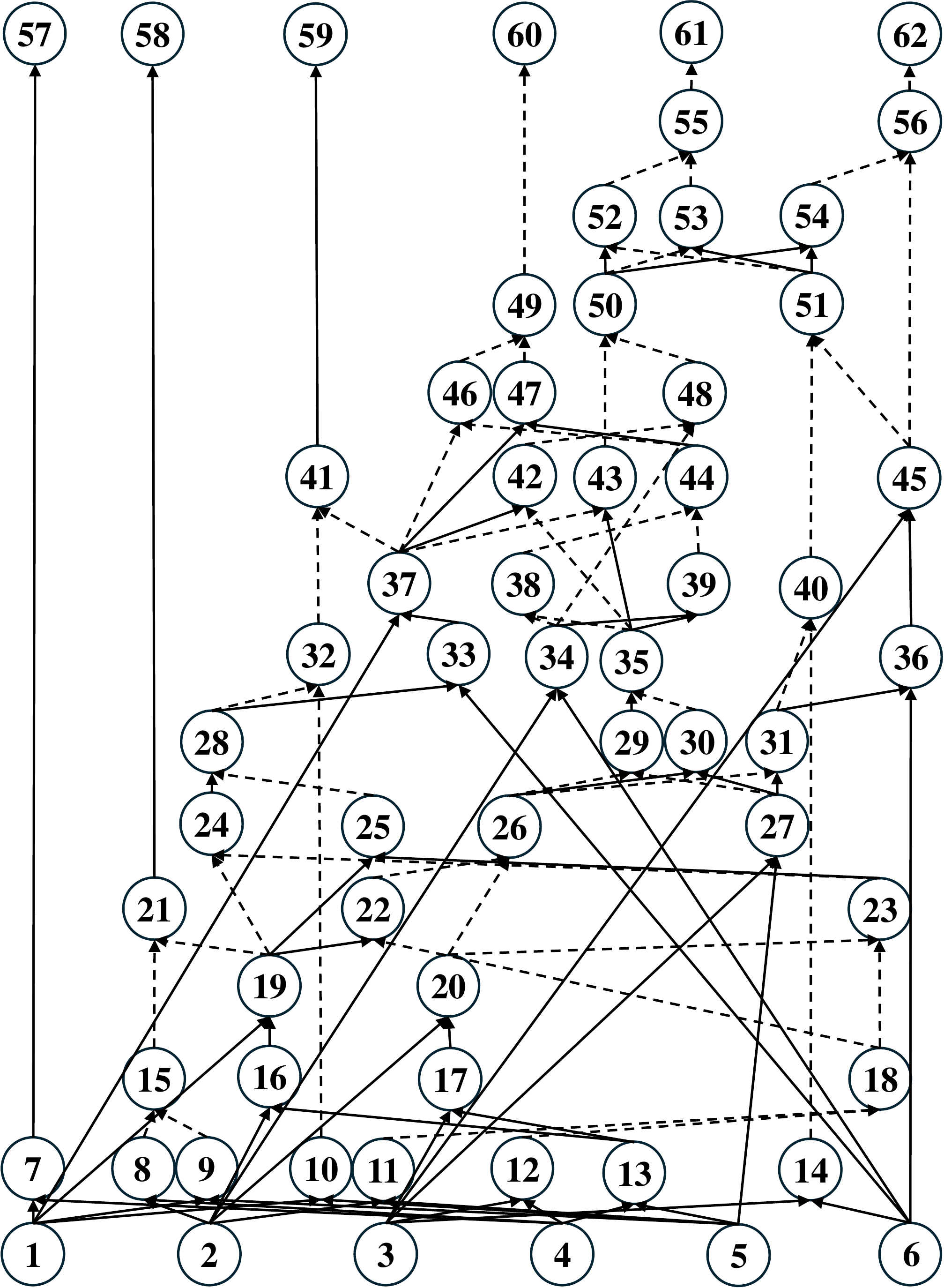}
        \caption{AIG of 3-bit CSA multiplier w 7nm ASAP mapping.}
        \label{fig:motivation_1}
    \end{subfigure}%
    \hspace{0.25em}
    \begin{subfigure}[t]{0.23\linewidth}
        \centering
        \includegraphics[width=\textwidth]{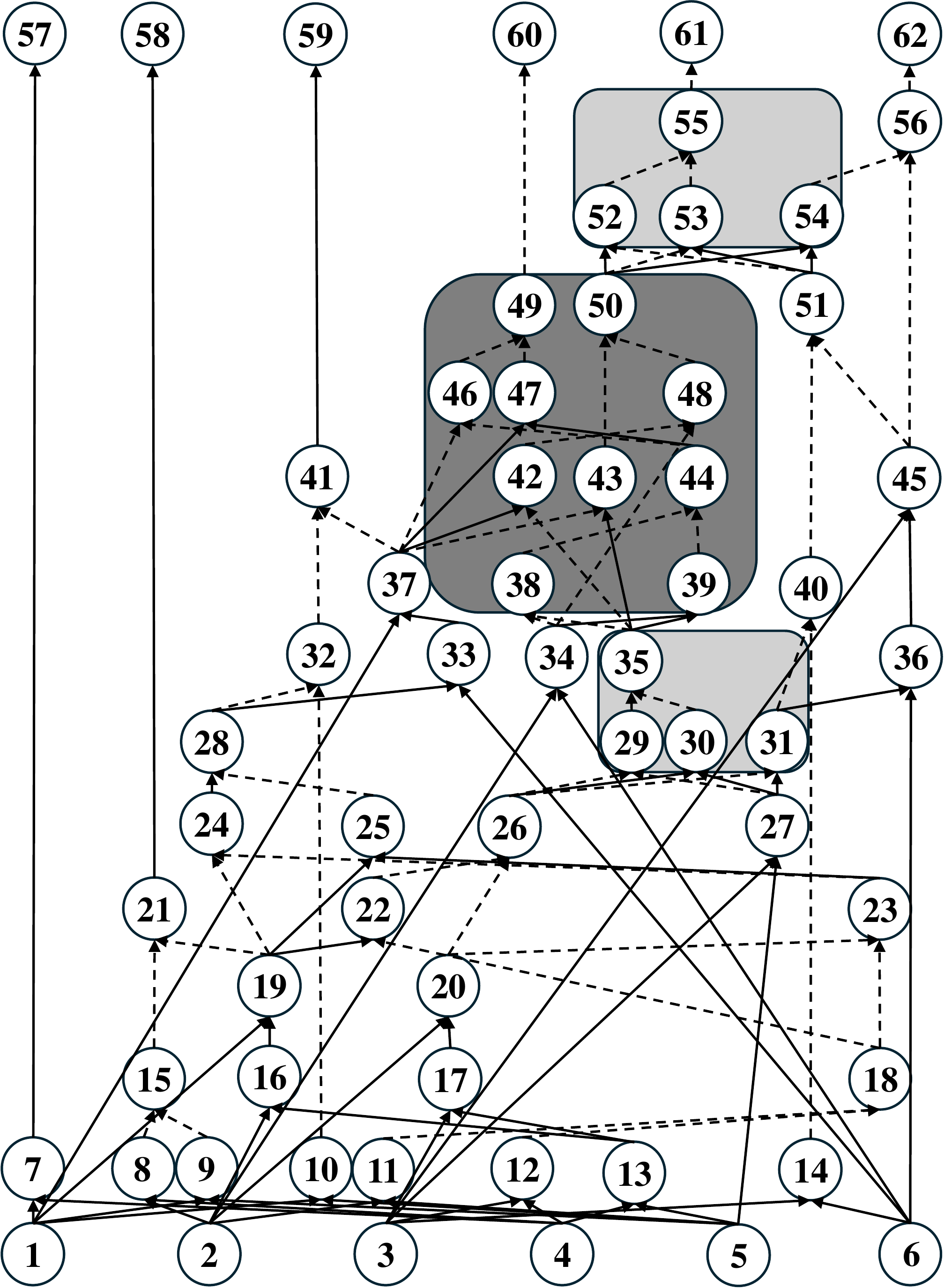}
        \caption{FA/HA block in the AIG.}
        \label{fig:motivation_2}
    \end{subfigure}%
    \hspace{0.25em}
    \begin{subfigure}[t]{0.23\linewidth}
        \centering
        \includegraphics[width=\textwidth]{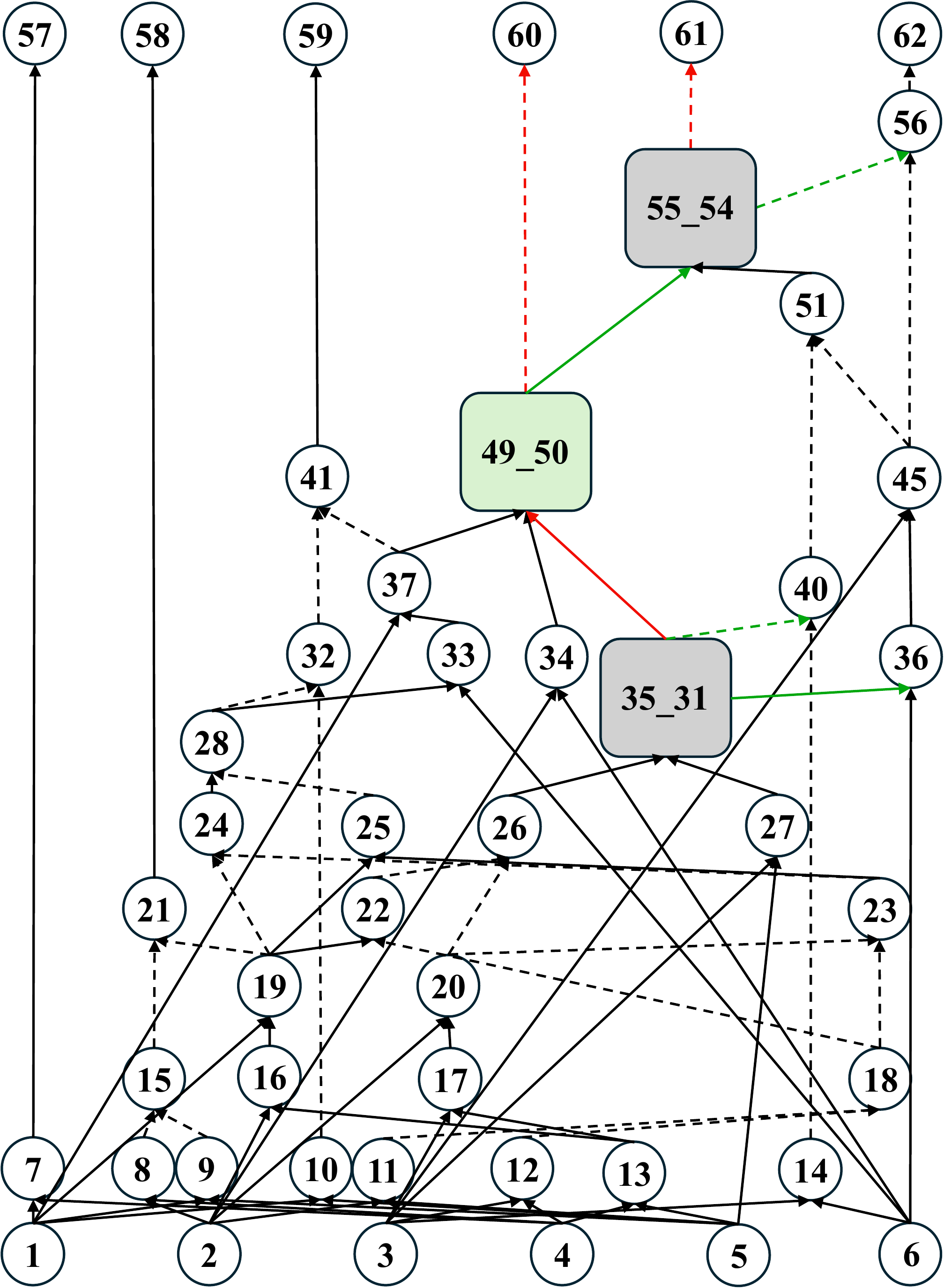}
        \caption{NPN FA block 49\_50 and HA blocks \{35\_31,55\_54\} by cut enumeration.}
        \label{fig:motivation_3}
    \end{subfigure}%
    \hspace{0.25em}
    \begin{subfigure}[t]{0.23\linewidth}
        \centering
        \includegraphics[width=\textwidth]{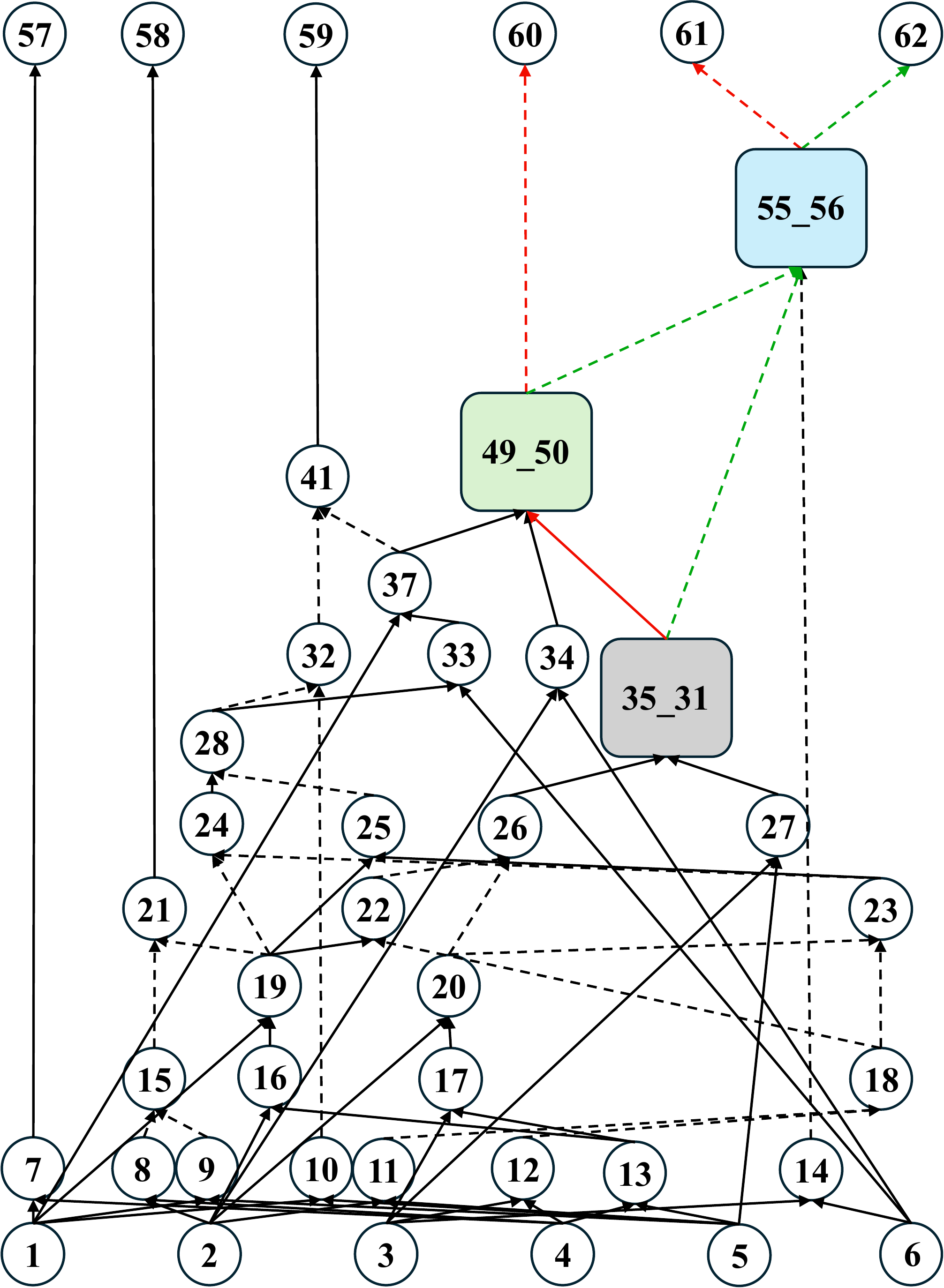}
        \caption{BoolE identifies an additional exact FA block 55\_56 is extracted.}
        \label{fig:motivation_4}
    \end{subfigure}
    
    \caption{Motivations for BoolE. The component blocks of exact FA, NPN FA, and HA are shown in blue, green, and grey blocks. The output signals of corresponding XOR and MAJ are marked as green and red correspondingly. (a) AIG of a 3-bit CSA multiplier generated by ABC and mapped using 7nm ASAP Technology Mapping. (b) FA and HA blocks in AIG detected by ABC. (c) Adder tree extracted from AIG. There is only one NPN FA. (d) The output netlist extracted after BoolE rewriting. An additional exact FA is reconstructed after BoolE rewriting.}
    \vspace{-3mm}
    \label{fig:motivations}
\end{figure*}

In this section, we present a motivating example in Figure~\ref{fig:motivations} to illustrate challenges in FA block identification. Figure \ref{fig:motivations}(\subref{fig:motivation_1}) shows the AIG of a 3-bit carry-save-array (CSA) multiplier after ASAP 7nm technology mapping \cite{xu2017standard}, with solid and dashed lines representing output signals and their negations, respectively.
Prior to technology mapping, the 3-bit CSA multiplier netlist contains 3 FAs. However, after technology mapping, ABC identifies only one NPN FA block through cut enumerations, as shown in Figure \ref{fig:motivations}(\subref{fig:motivation_2}). In this figure, the output of XOR and MAJ gates are highlighted in red and green, respectively.
Figure \ref{fig:motivations}(\subref{fig:motivation_3}) illustrates the adder tree of the 3-bit post-mapping CSA multiplier, where block \texttt{49\_50} is identified as an NPN FA (marked in green), and blocks \texttt{55\_54} and \texttt{35\_31} are HAs (marked in gray).


While NPN classification effectively recognizes structurally similar components, it does not guarantee logical equivalence to true FAs. For tasks such as formal verification, which require precise logical equivalence to ensure the correctness of a design, identifying NPN-equivalent FAs is insufficient. Formal verification demands that the functional behavior of the circuit matches the specification exactly, without deviations introduced in NPN transformations. For example, RevSCA-2.0\cite{mahzoon2021revsca} utilizes block identification of exact HAs and FAs to eliminate vanishing monomials by detecting and locally removing sources of redundant terms ahead of backward rewriting. However, this process requires exact logical equivalence to functional blocks.

Conventional Boolean reasoning methods, like cut enumeration, rely on static netlist structures to identify functional blocks such as FAs. However, processes like technology mapping and logic optimization often modify or eliminate these structures, rendering them unrecognizable to traditional tools. For the example in Figure \ref{fig:motivations}, only one FA can be identified with cut enumeration. In contrast, equality saturation rewrites netlists into functionally equivalent forms, enabling exhaustive exploration of alternative structures. This flexibility opens opportunities for reconstructing functional blocks, thereby improving the robustness and effectiveness of symbolic reasoning tasks.

Finally, we present the netlist extracted from BoolE in Figure~\ref{fig:motivations}(\subref{fig:motivation_4}), where BoolE has reconstructed the netlist structure for FA identification. Apart from the NPN FA \texttt{49\_50}, BoolE successfully identifies an additional exact FA, \texttt{55\_56} (marked in blue), after term rewriting. 
For a large-scale 128-bit technology-mapped CSA multiplier, BoolE reconstructs 14{,}713 out of 16{,}128 NPN FAs within the theoretical upper bound, which is 3{,}771 more than what ABC can identify. Furthermore, 3{,}418 of the reconstructed FAs are exact FAs, representing a $4.42\times$ performance improvement compared to ABC.

\section{Approach}\label{sec:approach}

\begin{figure*}[!htb]
    \centering
    \includegraphics[width=0.85\linewidth]{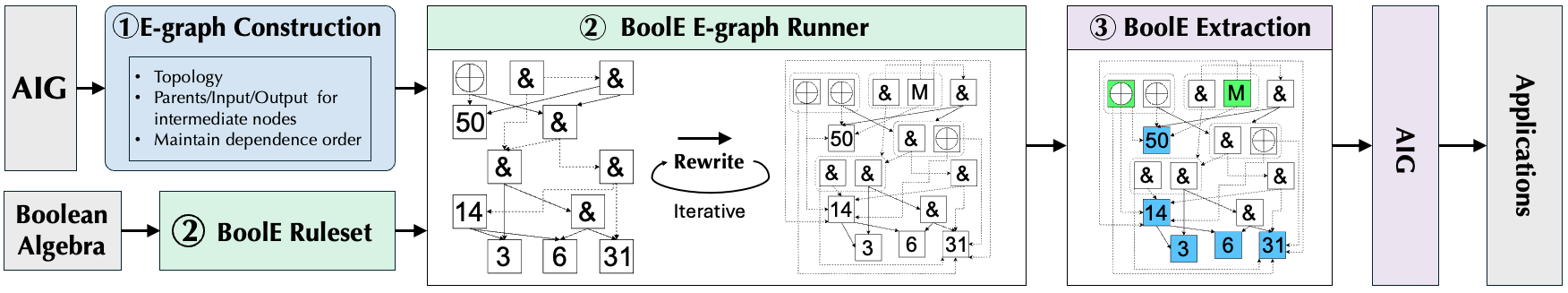}
    \vspace{-2mm}
    \caption{The overview of BoolE, an exact symbolic reasoning framework for Boolean netlists utilizing equality saturation. It enhances performance by employing a novel extraction algorithm specifically designed to identify and capture multi-input, multi-output exact FA structure. We incorporate part of the e-graph for motivating example shown in Figure \ref{fig:motivations} to illustrate how the additional extra FA is extracted from e-graph, where the extracted e-nodes are specially highlighted. The \texttt{XOR} and \texttt{MAJ} gates forming an exact FA are highlighted in green.}
    \vspace{-3mm}
    \label{fig:flow_egraph}
\end{figure*}

The overview of BoolE is illustrated in Figure \ref{fig:flow_egraph}. The framework accepts gate-level netlists as input, where each node represents a standard Boolean algebra operation (e.g., \texttt{AND}, \texttt{XOR}, \texttt{NOT}). Initially, BoolE parses the input AIG file, extracts necessary information, and converts the netlist into an e-graph. Subsequently, the BoolE rewriting engine applies rewriting rules to expand the e-graph, thereby identifying \texttt{MAJ} and \texttt{XOR} components. 
For e-graph extraction, we developed a DAG-based cost extraction algorithm to ensure the multi-input multi-output structure (i.e., FA) is correctly captured. Finally, BoolE converts the extracted e-graph into AIG 
format. And the output AIG can also be integrated into other tools (e.g., PolyCleaner and RevSCA\cite{mahzoon2018polycleaner,mahzoon2019revsca}) for further applications such as functional verification. 



\subsection{BoolE Methodology - Symbolic Reasoning via Equality Saturation}

To identify FAs and reconstruct adder tree structures, BoolE employs equality saturation to process the input netlists. Here, we introduce the core components of the framework: e-graph construction and rewriting ruleset.

\subsubsection{E-Graph Construction}
BoolE gathers functional information from input netlists to accurately represent the AIG structure and construct the e-graph based on this information. The algorithm for e-graph construction is detailed in Algorithm \ref{alg:egraph_construction}. Firstly, BoolE collects data on the primary inputs and primary outputs of the netlists. Furthermore, BoolE establishes the dependency relationships for each Boolean operation, defining the input and output edges for the nodes. Subsequently, BoolE incrementally inserts operation nodes into the e-graph following the topological order of the graph nodes. Specifically, operations are inserted into the e-graph from the leaf nodes to the root nodes to ensure that child nodes are processed beforehand. This topological insertion order is critical for accurately maintaining the dependencies between operations. Each insertion generates an identifier representing the e-nodes and connects these e-nodes with their respective input parameters. The e-graph construction is shown in part \circled{1} in Figure \ref{fig:flow_egraph}.

\algnewcommand{\Input}[1]{\Require{#1}}
\algnewcommand{\Output}[1]{\Ensure{#1}}
\algrenewcommand{\algorithmicrequire}{\textbf{Input:}}
\algrenewcommand{\algorithmicensure}{\textbf{Output:}}

\begin{algorithm}
\scriptsize
\caption{E-graph Construction}
\label{alg:egraph_construction}
\begin{algorithmic}[1]
   \Input{ Vertex list $V$ from parsing netlist} 
   \Output:{ e-graph $G_e$}

    \State Initialize $vmap$ as empty HashMap
    \For{each node $v$ in TopoSort($V$)} \Comment{From leaf to root}
    \State $in\_id \leftarrow \lbrack \rbrack$ 
    \For{each input $i$ for node $v$}
    \State $in\_id$.pushback($vmap$[$i$]) \Comment{All children are inserted}
    \EndFor
    \State $id \leftarrow $ $G_e$.insert($v, in\_id$) \Comment{Insert $v$ to e-graph}
    \State $vmap[v] = id$
    \EndFor

\end{algorithmic}
\end{algorithm}


   



\subsubsection{E-Graph Rewriting}

\label{sec:rewriting}

The rulesets are composed of two parts: (1) basic Boolean rules (\( R_1 \)), such as the commutative law, associative law, and De Morgan's Laws, which aim to expand the e-graph by saturating it with more functionally equivalent expression trees; (2) rules for \texttt{XOR} and \texttt{MAJ} identification (\( R_2 \)), which directly identify \texttt{XOR} and \texttt{MAJ} operations within the e-graph. A subset of these rewriting rules is presented in Table \ref{tbl:rewriting_rule}. In total, we collected 68 rules in \( R_1 \) and 119 rewriting rules for \( R2 \). To construct \( R_2 \), we utilized 8- to 128-bit Booth and CSA multipliers as templates and employed the synthesis tool ABC to detect the adder trees within the netlists and extract the structural patterns of \texttt{MAJ} and \texttt{XOR} operations. We constructed the corresponding e-graph rewriting rules $R_2$ and eliminated duplicate rules from the rule set. 

The ruleset construction and e-graph rewriting correspond to part \circled{2} in Figure \ref{fig:flow_egraph}. Here we introduce 3 optimization tricks we employ in e-graph rewriting:

\begin{enumerate}
    \item To enhance scalability while keeping performance, BoolE offers a lightweight version of rewriting rules. This version is carefully designed through empirical, manual pruning to ensure effectiveness for large-scale benchmarks, striking a balance between reducing complexity and maintaining reasoning performance.
    \item Additionally, the two parts of the rewriting rules are applied incrementally in two phases. Firstly, we first apply \( R_1 \) to the initial e-graph with 10 iterations of rewriting. Subsequently, based on the \( R_1 \) rewrited e-graph, we apply \( R_2 \) with 3 iterations. This incremental saturation approach allows fine adjustment of the iteration limits for \( R_1 \) and \( R_2 \) separately.
    \item After e-graph saturation, BoolE deletes redundant e-nodes that do not contribute to performance improvements. This reduces unnecessary memory usage and runtime. For instance, based on the commutative property, expressions such as $\texttt{XOR}(a, b, c)$, $\texttt{XOR}(b, a, c)$ belong to the same e-class. 
    BoolE retains only one unique expression within each e-class, deleting the other equivalent e-nodes. A similar approach is applied to $\texttt{MAJ}(a, b, c)$ and FA$(a, b, c)$ expressions.

\end{enumerate}

\begin{table}[!ht]
    \centering
    \renewcommand{\arraystretch}{1.25} 
    \scriptsize
     \caption{Example BoolE rewriting rules. The complete rewriting library consists of 68 Boolean rewriting rules and 39/90 additional rules specifically designed for the identification of \texttt{MAJ}/\texttt{XOR} operations.}
    \label{tbl:rewriting_rule}
    \begin{tabular}{p{1.09cm} p{5.5cm} p{1.3cm}}
    \hline
        ~ & Pattern & Transformation \\ \hline
        Basic rules & $a \, \& \, b$ & $ b \, \& \, a$ \\ \hline
        ~ & $a$ & $(\,a\,')\,'$ \\ \hline
        ~ & $(a\,\&\,b)'$ & $a'\,||\,b'$ \\ \hline
        ~ & $(a\,||\,b)'$ & $a'\,\&\,b'$ \\ \hline
        ~ & $(a\,\&\,b)\,\&\,c$ & $a\,\&\,(b\,\&\,c)$ \\ \hline
        MAJ rules & $(a \, \& \, b) \, || \, (a \, \& \, c) \, || \, (b \, \& \, c)$& $\text{\texttt{MAJ}}(a,b,c)$ \\ \hline
        ~ & $(a' \, \& \, (b \, \& \, c)')' \, \& \, (b' \, \& \, c')'$ & $ \text{\texttt{MAJ}}(a,b,c)$ \\ \hline 
        XOR rules & $(a  \&  b' \&  c') \, ||  (a' \&  b  \&  c')  ||  (a'  \&  b'  \&  c)  ||  (a  \&  b  \&  c)$ & $ \text{\texttt{XOR}}(a,b,c)$ \\ \hline
        ~ & $ \left(\left( \left( a  ||  (b  \&  c) \right)  || (b  ||  c)' \right) \& \left( a'  || \left( (b \&  c)'  \&  (b  ||  c) \right) \right)\right)' $
 & $ \text{\texttt{XOR}}(a,b,c)$ \\ \hline
    \end{tabular}
\end{table}

\subsection{BoolE E-Graph Exact Extraction}

In this section, we introduce our e-graph extraction for multi-input multi-output FA structure, which corresponds to part \circled{3} in Figure \ref{fig:flow_egraph}.

\textbf{Cost Function}: Our cost function aims to maximize the number of exact FAs in the extracted expression within the e-graph \( G \). Specifically, for an expression tree \( t_e \) composed of e-nodes \( e \), the cost function \( C(t_e) \) is defined as: $C(t_e) = \sum_{e \in t_e} -1_{\{\text{e is exact FA}\}}$. 
E-graph extraction seeks the optimal expression tree \( t^* \) with the lowest total cost from \( \mathcal{T} \), the set of all possible expression trees in \( G \):

\[
t^* = \arg\min_{t_e \in \mathcal{T}} C(t_e)
\]
Each expression tree \( t \) is constructed by selecting one node from each e-class, ensuring conformity to the e-graph structure.

\textbf{Multi-output FA structure}: Standard e-graph structures typically employ prefix notation expressions, assuming that all operators have single outputs. Consequently, the e-graph framework inherently lacks support for multi-output operations, only focusing on single-output functionalities. However, FA
is a multi-input, multi-output structure, consisting of three inputs and two outputs representing the carry and sum, respectively.

\begin{figure}[!htb]
    \centering
    \includegraphics[width=0.6\linewidth]{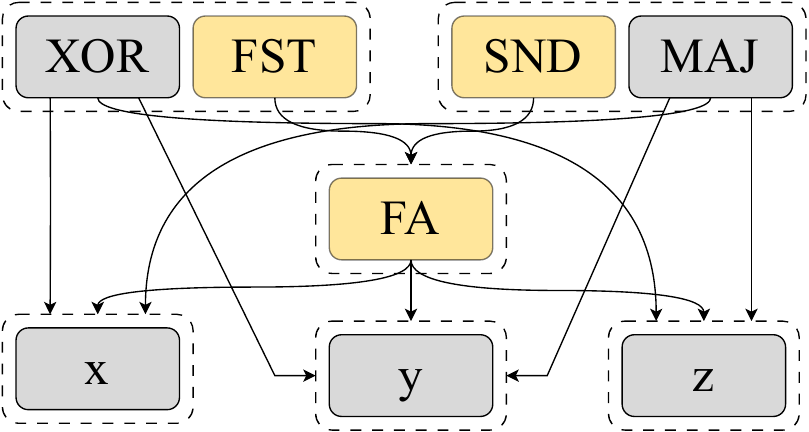}
    \vspace{-2mm}
    \caption{FA Structure: \texttt{XOR} and \texttt{MAJ} e-nodes with the exactly same inputs will be paired. Each paired \texttt{XOR}/\texttt{MAJ} operation will insert a FA nodes and utilize \textit{FST/SND} to project out the carry/sum bits. \textit{FST/SND} will unified to the e-classes of \texttt{XOR}/\texttt{MAJ}}
    \label{fig:fa_struct}
\end{figure}

To accommodate multi-output operations, equality saturation introduces new e-nodes into the e-graph. BoolE iterates through all e-nodes, identifying \texttt{XOR} and \texttt{MAJ} nodes with the exact same inputs. When such nodes are found, BoolE inserts an FA node that shares the same inputs and generates a tuple of outputs for carry and sum. Subsequently, BoolE employs the \texttt{FST} and \texttt{SND} pseudo-operations to project the individual carry and sum signals from the output tuple. Finally, BoolE unifies the e-classes of the \texttt{FST} and \texttt{SND} operations with the corresponding \texttt{XOR} and \texttt{MAJ} nodes, ensuring their equivalence within the e-graph framework. The structure of the FA is depicted in Figure \ref{fig:fa_struct}. It is crucial to note that extracting FA structures is treated as an atomic operation. Thus, \texttt{FST}, \texttt{SND}, and FA must be selected collectively or not at all. This atomic extraction ensures the integrity of the FA structure is maintained during BoolE extraction.

\textbf{DAG based extraction algorithm}: BoolE utilizes DAG cost extraction, which counts each shared sub-expression only once for greater accuracy, to prevent double-counting of FA numbers. The extraction process is detailed in Algorithm~\ref{alg:extraction}. The algorithm first initializes a queue with all leaf nodes and an empty cost map for each e-class. It then iteratively processes nodes by dequeuing them and checking if all child nodes have computed costs. If so, it calculates the new cost for the node. If the new cost is lower than the previous cost, the cost map is updated, and parent nodes are enqueued. Nodes labeled \texttt{FST}/\texttt{SND}, and FA are handled specially in a post-processing step to ensure the integrity of FA structures. Each e-class maintains a cost map with the current best cost, selected e-node, and chosen child e-classes. By iteratively updating costs, the algorithm efficiently constructs the optimal expression tree \( t^* \). This hash map, mapping child e-class IDs to their corresponding costs, introduces memory usage issues as the input e-graph grows. To save memory, BoolE uses flexible data types for the keys and values in the cost hash map. For small e-graphs (fewer than 65{,}536 e-classes), BoolE uses unsigned 16-bit integers for hash map keys, conserving memory. For larger e-graphs, it adjusts to unsigned 32-bit integers. This adaptability based on e-graph size ensures efficient memory utilization.

\begin{algorithm}
\caption{E-Graph Extraction}
\scriptsize
\label{alg:extraction}
\begin{algorithmic}[1]
   \Statex \textbf{Input:} e-graph $G_e$ \hspace{1cm} \textbf{Output:} Optimal $t^*$ w.r.t. $Cost(G_e)$

    \Function{CalculateCostSet}{$G_e$, node, Costs}
      \State $\text{results} \leftarrow \text{\{key:Costs[key].results}\}$ for key in node.children()
      \State total $\leftarrow$ Sum(result.values())
      \State \Return \{results,total,node\}
    \EndFunction

    \Function{Extract}{$G_e$, root}
    \State $Queue$ $\leftarrow$ LeafNode($G_e$) \Comment{Queue of e-nodes to be processed}
    \State $t^*=\emptyset$ \Comment{Expression tree to be constructed}
    \State $Costset=\emptyset$ \Comment{Initial empty map of e-class to cost}
    \While{$Queue$ is not empty}
      \State node = $Queue$.pop()
      \If{all c in node.children() are in $Costset$}
      \State prev\_cost = $Costset$[class\_id].total
      \State new\_cost = CalculateCostSet($G_e$, node, $Costset$)
      \If{new\_cost.total $<$ prev\_cost}
      \State $Queue$.insert(node.parents()) \Comment{Cost map update}
      \State $Costset$[node]= new\_cost
      \If{node is FST/SND}
      \State PostProcessing() \Comment{FA structure integrity check}
      \EndIf
      \EndIf
      \EndIf
    \EndWhile
    \EndFunction

\end{algorithmic}
\end{algorithm}

\section{EXPERIMENTAL RESULTS}\label{sec:results}




The experiments were conducted on a system with an Intel(R) Xeon(R) Gold 6418H CPU, featuring 48 physical cores and 2 threads per core, and 1 TByte of RAM. We use the \textit{egg} framework~\cite{willsey2021egg} as the back-end to develop \textbf{BoolE} for e-graph construction and term rewriting. The experimental benchmarks include unsigned CSA multipliers and signed Booth multipliers, collected from the state-of-the-art (SOTA) baseline works ABC \cite{yu2017fast} and Gamora \cite{wu2023gamora}. Specifically, the AIG-based approach (\texttt{\&atree}) in ABC leverages structural hashing and functional pattern matching that reconstructs the adder tree structure via the cut enumeration algorithm \cite{yu2017fast}. Gamora is a SOTA graph learning-based Boolean reasoning tool \cite{wu2023gamora} that targets the same objective, while the training dataset is collected with the AIG-based approach \cite{yu2017fast}.





\subsection{Performance of Symbolic Reasoning}

To demonstrate the performance of BoolE, we prepared three research questions(RQs) in this section:

\noindent
\textbf{RQ1: How effective of BoolE without structure rewriting?} -- We collected the number of FAs identified by ABC and BoolE across various bitwidths for pre-mapping netlists. 
All adder tree structures are present in the netlists before technology mapping or logic optimization.
For instance, in an \( n \)-bit CSA multiplier, the theoretical upper bound of FA number is \( (n-1)^2 - 1 \), which can be determined through exhaustive cut enumeration.

\begin{figure*}[htbp]
    \hfill
    \begin{subfigure}[t]{0.8\linewidth}
        \hfill
        \includegraphics[width=\textwidth]{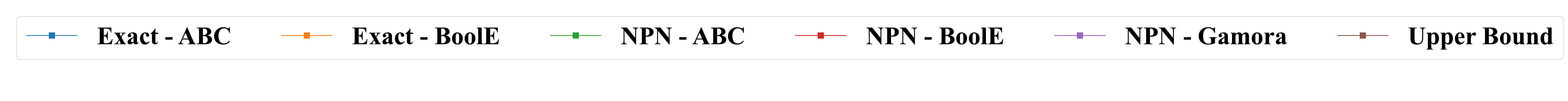}
        \label{fig:fa_csa}
    \end{subfigure}%
    \vspace{-8mm}
    \hfill
    \begin{subfigure}[t]{0.46\linewidth}
        \centering
        \includegraphics[width=\textwidth]{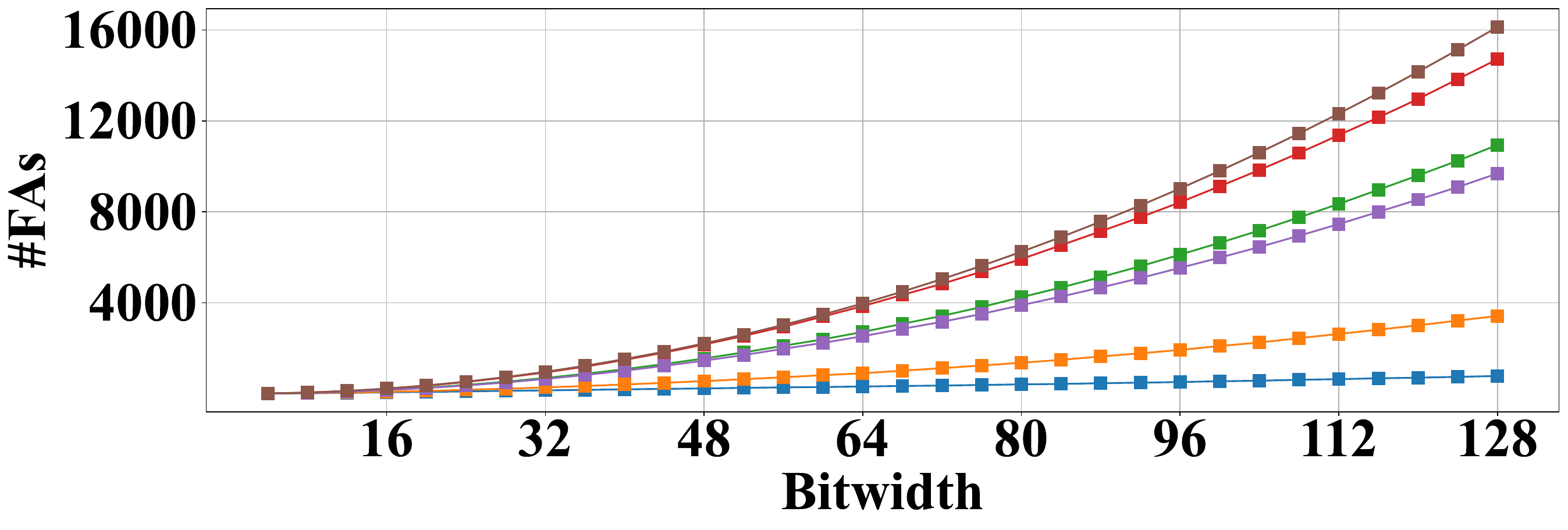}
        \label{fig:fa_csa}
    \end{subfigure}%
    \hfill
    \begin{subfigure}[t]{0.46\linewidth}
        \centering
        \includegraphics[width=\textwidth]{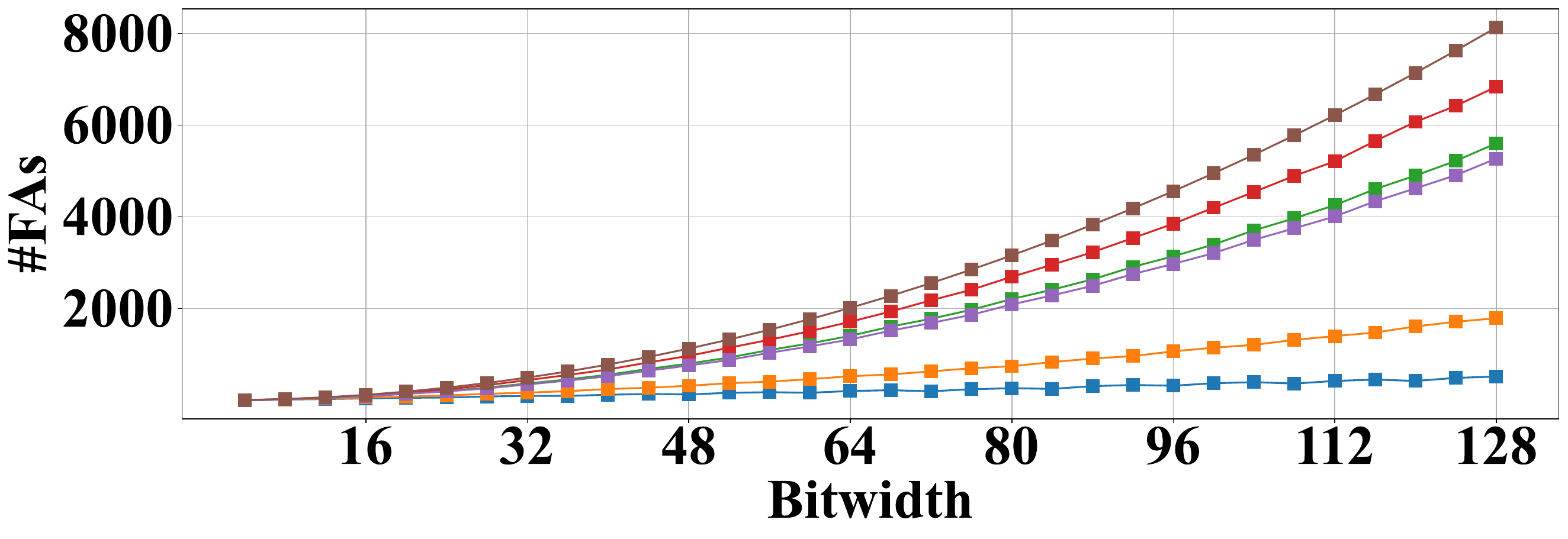}
        \label{fig:fa_booth}
    \end{subfigure}
    \vspace{-3mm}
    \caption{Performance comparison among BoolE, ABC, and Gamora for CSA (left) and Booth (right) multipliers after ASAP 7nm technology mapping. BoolE consistently outperforms ABC and Gamora, identifying $3.53\times$ and $3.01\times$ exact FAs than ABC in CSA and Booth multipliers. And BoolE achieves an average of $93.48\%$ and $84.81\%$ of the theoretical upper bound in number of NPN FAs, respectively. 
    }
    \vspace{-4mm}
    \label{fig:result1}
\end{figure*}

ABC can identify all adder trees in pre-mapping netlists successfully.
Figure \ref{fig:result1} presents the results alongside the theoretical upper bound for the number of pre-mapping FAs. In this figure, the upper bound curve data also represents the number of pre-mapping FAs identified by ABC and BoolE.
\textbf{And BoolE also achieves the optimal solution for both CSA and Booth benchmarks across all bitwidths for pre-mapping benchmarks.}
Since all NPN FAs can be exhaustively identified in pre-mapping netlists without rewriting, the reasoning performance is entirely dominated by ruleset \( R_2 \) described in Section~\ref{sec:rewriting}, which is specialized for \texttt{XOR} and \texttt{MAJ} identification.
The result underscores the superior performance of ruleset \( R_2 \), which dominates the reasoning capability to discover all existing NPN FAs in pre-mapping netlists.

\noindent
\textbf{RQ2: How effective is BoolE under heavily logic optimization and technology mapping?} -- Reasoning high-level components under heavily logic optimization and technology mapping settings is known to be significant more challenging \cite{wu2023gamora}\cite{yu2017fast}\cite{deng2024less}\cite{li2013wordrev}, which is more critical in real-world datapath synthesis \cite{coward2024combining, yu2016automatic} and formal verification scenarios \cite{yu2017fast, yu2016formal, mahzoon2021revsca}.
Therefore, our objective is to demonstrate the robustness of BoolE under technology mapping and logic optimization,
which reflects the rewriting and extraction capability of BoolE.

To evaluate the performance of BoolE on technology-mapped netlists, we quantified the number of FAs discovered by BoolE across technology-mapped CSA and Booth multipliers of various bitwidths.
We utilized the synthesis tool ABC to perform technology mapping with the 7nm ASAP library, which encompasses a diverse set of 161 standard-cell gates. 
The test results are presented in Figure \ref{fig:result1}, with the CSA and Booth multiplier results shown in the left and right subfigures, respectively. 
The x-axis represents the bitwidth of the benchmark multipliers, while the y-axis indicates the number of FAs discovered using different reasoning tools: BoolE, ABC, and Gamora.

The results demonstrate that BoolE successfully reconstructs the majority of FA structures, and provides superior exact symbolic reasoning performance for multipliers compared to ABC and Gamora. Later in Section \ref{sec:verification}, we also demonstrate its robustness against formal verification tool RevSCA-2.0, which integrates the same reasoning function.
\textbf{Specifically, BoolE achieves an average of $93.48\%$ and $84.81\%$ of the theoretical upper bound for NPN FAs in technology-mapped CSA and Booth multipliers, respectively.}
In comparison, ABC can only identify $68.07\%$ and $69.28\%$ of the maximum NPN FAs for CSA and Booth multipliers on average, respectively.
Furthermore, Gamora reasoning performance drops to $63.81\%$ and $66.68\%$ for CSA and Booth multipliers, respectively. BoolE constantly outperforms ABC and Gamora for post-mapping netlists.
Additionally, BoolE focuses on reconstructing as many exact FAs as possible. \textbf{The number of exact FAs after BoolE rewriting increased by $3.53\times$ and $3.01\times$ compared with ABC for CSA and Booth multipliers, respectively.} 

\begin{figure}[!htb]
    \centering
    \includegraphics[width=\linewidth]{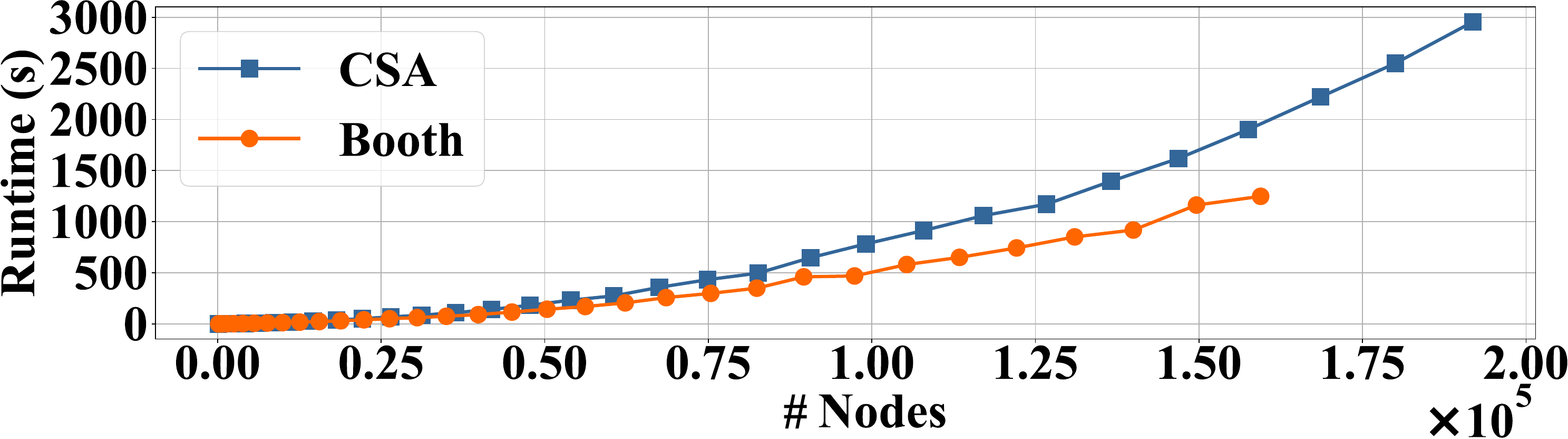}
    \vspace{-4mm}
    \caption{BoolE runtime of post-mapping CSA and Booth multiplier w.r.t. the size of the input netlist (AIG node number).}
    \label{fig:runtime}
    \vspace{-3mm}
\end{figure}

\noindent
\textbf{RQ3: Is BoolE scalable?} -- To demonstrate the scalability of BoolE, we present the end-to-end runtime across all tested benchmarks. Specifically, we recorded the runtime and number of nodes in the netlist of each benchmark, including post-mapping CSA and Booth multipliers. The results are illustrated in Figure~\ref{fig:runtime}.





Although functional rewriting for Boolean networks involves higher time complexity compared to simple structural matching in cut enumeration, BoolE achieves efficient rewriting runtimes. It completes all benchmarks for post-mapping Booth and CSA multipliers from 4-bit to 128-bit within 50 minutes. For the largest case, 128-bit post-mapping CSA multipliers containing 191{,}863 AIG nodes, the rewriting process takes 50 minutes. The runtime for BoolE rewriting is insignificant compared to the order-of-magnitude improvements it brings to real-world applications. In tasks such as formal verification, the reasoning performance directly impacts the overall complexity and runtime. BoolE's superior reasoning performance enables orders of magnitude speedup in the formal verification of larger, optimized multipliers. For example, BoolE takes only 11 seconds to reason the netlist for a 24-bit multiplier, while the verification runtime is drastically reduced from 32{,}402.74 seconds to just 0.07 seconds. The runtime spent on the rewriting process is well compensated by the substantial reductions achieved in end-to-end verification time. The runtime and efficiency for formal verification tasks will be elaborated in Section \ref{sec:verification}.

\vspace{-1mm}
\subsection{Integrated Application to Formal Verification}\label{sec:verification}

\textbf{RQ4: Can BoolE practically enhance real-world applications orthogonally (formal verification)?} 
Boolean symbolic reasoning is a critical topic in various domains, such as functional verification and datapath optimization.
We take verification of multipliers as a case study to illustrate how BoolE can enhance real-world applications. Boolean symbolic reasoning, as an orthogonal technique, can be integrated into many formal verification tools for reverse engineering\cite{mahzoon2019revsca,mahzoon2021revsca,ritirc2018improving}. Here we utilize RevSCA-2.0 \cite{mahzoon2021revsca} as our back-end tool. The complexity of symbolic computer algebra-based multiplier verification is determined by the maximum polynomial size (\# monomials) \cite{yu2016formal}, and its performance is measured by verification runtime.
To mitigate polynomial explosion, RevSCA-2.0 uses cut enumeration to detect functional blocks, including exact FAs. It then eliminates vanishing monomials within these blocks before backward rewriting. Consequently, reasoning performance directly impacts the detection of functional blocks, the elimination of vanishing monomials, and the overall success of verification.

ABC offers a logic circuit optimization feature that simplifies Boolean expressions and reduces gate counts. This optimization functionality is integrated into the \textit{dch} interface.
We evaluated the total runtime of verifying a CSA multiplier benchmark, optimized using \textit{dch} bit-optimization, under two configurations: with and without BoolE optimization. Note that RevSCA-2.0 is equipped with a functional reasoning engine with cut enumeration. The test results are presented in Table \ref{table:result}. This table provides the verification runtime across various bitwidths. Additionally, we report the maximum polynomial size (Max Poly Size) observed during the backward iteration process, which reflects the complexity of the verification tasks. It is important to note that we only report the number of exact FAs because NPN FAs cannot be detected as functional blocks, and only exact FAs are beneficial for multiplier verification simplification. Verification runtime exceeding 72 hours is marked as time out (TO).

We can see a significant improvement in runtime brought by BoolE rewriting. The formal verification of CSA multipliers after logic optimization gets timed-out without BoolE rewriting when the bitwidth is larger than \( 24 \) bits. After BoolE rewriting, the multipliers of up to \( 128 \) bit can be formally verified. And there is a significant speedup in verifying multipliers with higher bitwidths. 
For example, BoolE provides $2825\times$ runtime improvement for 24-bit \textit{dch}-optimized multiplier via exact symbolic reasoning. 
Furthermore, BoolE enables the verification of \textit{dch}-optimized multipliers with bitwidths exceeding 24 bits. However, without BoolE, RevSCA-2.0 cannot formally verify these larger multipliers within a \textbf{72-hour timeout}.

As illustrated in the fourth column of Table \ref{table:result}, the number of exact FAs detected by ABC in CSA multipliers drops to zero due to logic optimization. This disappearance of exact FAs significantly increases the complexity of the verification process during backward rewriting, especially as the bitwidth scales. On the other hand, BoolE can reconstruct most disappeared functional blocks. In the third column of Table \ref{table:result}, we present the number of exact FAs reconstructed by BoolE, which is compared with the upper bound in the second column which represents the maximum number of FAs in CSA multipliers. BoolE successfully reconstructs up to 99.2\% of the exact FAs within the theoretical upper bound.



\begin{table}[t]
    \centering
    \vspace{0mm}
    \caption{Verification results of \texttt{dch}-optimized CSA multipliers using RevSCA-2.0 under two configurations: with BoolE optimization (BoolE) and without optimization (Baseline).}
    \resizebox{0.5\textwidth}{!}{%
        \normalsize
        \begin{tabular}{|c|c|c|c|c|c|c|c|}
            \hline
            \multirow{2}{*}{Bitwidth} & \multicolumn{3}{c|}{Number of Exact FAs} & \multicolumn{2}{c|}{Max Poly Size} & \multicolumn{2}{c|}{End-to End Runtime (s)} \\
            \cline{2-8}
            & Upper Bound & BoolE & Baseline & BoolE & Baseline & BoolE & Baseline \\
            \hline
            8 & 48 & 41 & 0 & 85 & 373 & 0.445 & 0.035 \\
            12 & 120 & 109 & 0 & 173 & 2173 & 0.955 & 0.297 \\
            16 & 224 & 209 & 0 & 293 & 34345 & 2.474 & 17.332 \\
            20 & 360 & 341 & 0 & 445 & 548917 & 5.165 & 1471.574 \\
            24 & 528 & 505 & 0 & 629 & 8781889 & 11.470 & 32402.739 \\
            28 & 728 & 701 & 0 & 845 & - & 22.951 & TO \\
            32 & 960 & 929 & 0 & 1093 & - & 43.910 & TO \\
            64 & 3968 & 3905 & 0 & 4229 & - & 1232.219 & TO \\
            96 & 9024 & 8929 & 0 & 9413 & - & 8748.443 & TO \\
            128 & 16128 & 16001 & 0 & 16645 & - & 33255.292 & TO \\
            \hline
        \end{tabular}
    }
    \label{table:result}
    \vspace{-5mm}
\end{table}

\vspace{-2mm}
\section{Conclusion}
\vspace{-1mm}
In conclusion, this paper introduces BoolE, an exact Boolean symbolic reasoning framework that leverages equality saturation techniques in e-graphs to overcome the limitations of conventional methods. By assembling a comprehensive ruleset and implementing careful optimizations, BoolE achieves enhanced performance in Boolean symbolic reasoning, enabling accurate Boolean function detection. BoolE effectively reconstructs multi-output structures, such as FAs, within technology-mapped and optimized netlists. Our evaluations demonstrate BoolE's significant improvements, highlighting its potential to advance digital circuit synthesis and verification.

\textbf{Acknowledgment} -- This work is supported by National Science Foundation (NSF) under CCF2350186, CCF2403134, CCF2349670, and CCF2349461 awards.


\small
\bibliographystyle{IEEEtranS}
\bibliography{ref.bib,cds.bib}

\end{document}